\documentclass[aps,prl,reprint,superscriptaddress]{revtex4-1}
\usepackage{pdfpages}
\usepackage{amsmath,amssymb}
\usepackage{bm}
\usepackage{graphicx}
\usepackage{hyperref}
\usepackage{tabularx}
\usepackage{subfigure}

\newcolumntype{C}{>{\centering\arraybackslash}X}

\makeatletter
\AtBeginDocument{\let\LS@rot\@undefined}
\makeatother

\begin{document}
\title{Ground state degeneracy in quantum spin systems protected by crystal symmetries}
\author{Yang Qi}
\affiliation{Department of physics, Massachusetts Institute of Technology, Cambridge, MA 02139, USA}
\author{Chen Fang}
\affiliation{Institute of Physics, Chinese Academy of Sciences, Beijing 100190, China}
\author{Liang Fu}
\affiliation{Department of physics, Massachusetts Institute of Technology, Cambridge, MA 02139, USA}
\begin{abstract}
We prove a theorem on the ground state degeneracy in quantum spin systems on two-dimensional lattices: if a half-integer spin is located at a center of symmetry where the point group symmetry is $\mathbb D_{2,4,6}$, there must be a ground state degeneracy. The presence of such  degeneracy in the thermodynamic limit indicates either a broken-symmetry state or a unconventional state of matter. Compared to the Lieb-Schultz-Mattis theorem, our criterion for ground state degeneracy is based on  the spin at each center of symmetry, instead of the total spin per unit cell. Therefore, our result is even applicable to certain systems with an even number of half-integer spins per unit cell.
\end{abstract}
\date{\today}
\maketitle


For quantum many-body systems with an odd number of spin-$\frac12$ per unit cell, the Lieb-Schultz-Mattis (LSM) theorem and its generalization to higher dimensions~\cite{LSM,Affleck1986,Affleck1987,Yamanaka1997,OshikawaLSM2000,Hastings2004} guarantee a ground-state degeneracy protected by the translation symmetry and the spin-rotation symmetry. Such a ground-state degeneracy rules out the possibility of a featureless paramagnetic phase, and indicates either a broken-symmetry state or a unconventional state of matter, such as a quantum spin liquid ~\cite{BalentsSLReview,SavarySL2016,ZhouSL2017} with topological order\cite{WenCSL1989,WenTO1990}.
Recently, LSM-type theorems have been developed for systems with both time-reversal symmetry and (magnetic) space-group symmetry~\cite{Parameswaran2013,Roy2012X,Watanabe2015,PoLSM2017,YMLuLSM,YangDyonLSM}.

In this work, we present a new theorem on ground state degeneracy in quantum spin systems, which solely relies on crystal symmetries, and specifically, the point groups.
Our theorem states that quantum spin systems in two-dimensional (2D) lattices where a half-integer spin is located at a center of symmetry with the point group $\mathbb D_{n}$ for $n=2,4$ or $6$~\footnote{For simplicity, in this paper we consider a strictly 2D lattice with a wallpaper group symmetry, and do not distinguish the symmetry groups $\mathbb D_n$ and $C_{nv}$. In general, our theorem applies to both cases.}, must have a ground state degeneracy.

Several remarks are in order: 1. Here and throughout this paper, a half-integer (integer) spin on a given site  refers to the spin degrees of freedom arising from an odd (even) number of electrons localized at the site. Importantly, our theorem does not rely on the presence of full spin-rotation symmetry, hence is applicable to systems with spin-orbit coupling. 2. In contrast with the original LSM theorem and its recent generalizations, our theorem does not involve any internal symmetry such as time-reversal. 3. Our theorem is applicable to a number of systems with an even number of half-integer spins in the unit cell.

For concreteness, we first derive this theorem for a rectangular lattice with wallpaper group p2mm, before presenting the generalization to other 2D lattices. Finally we discuss possible applications of our theorem to real materials and its possible generalizations. Some mathematical details of the proof of the theorem is provided in the Supplemental Material.

\paragraph{Rectangular lattice.}
\label{sec:rect}

Consider quantum spin systems on a rectangular lattice with the 2D wallpaper group $G=\text{p2mm}$. Any Hamiltonian satisfying the symmetry $G$, with or without spin-orbit coupling, must be invariant under the the action of any crystal symmetry operation $R \in G$ on  the lattice and on the spins jointly. This action is represented by a unitary transformation on many-body basis states:
\begin{equation}
R: \prod_j | s_j \rangle \rightarrow \prod_{j} U_j(R) | s_{j'=Rj} \rangle
\end{equation}
where $|s_j\rangle$ denotes the spin state on site $j$, $R$ maps site $j$ to $j'=Rj$, and $U_j(R)$ represents the action of $R$ on the spin state on site $j$. For translationally invariant systems, the operators $U_j(R)$ on different sites connected by primitive lattice vectors are identical.

Let us now consider a subgroup of $G$ that leaves the center of the unit cell (denoted by $a$) invariant (or the point group at $a$), denoted by $G_a$. $G_a = \mathbb D_2=\mathbb Z_2\times\mathbb Z_2$ is a group generated by the $C_2$ rotation and a mirror reflection. Since the Hamiltonian considered here is invariant under $G_a$, every energy eigenstate must belong to a certain representation of $G_a$. 

\begin{figure}
  \includegraphics{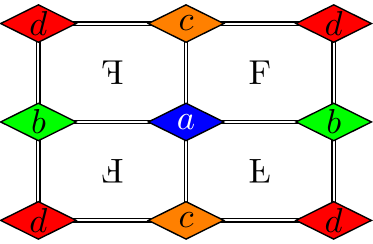}
  \caption{The symmetry elements of the p2mm group (rectangular lattice) in a unit cell. The double lines are mirror axes, and the rhombus symbols are centers of $\mathbb D_2$ symmetries. There are four inequivalent centers of symmetry, labeled as $a$-$d$, distinguished by different colors and orientations of the rhombuses. The letters F indicates symmetry-permitted degrees of freedom that can be decorated onto the lattice.}
  \label{fig:p2mm:sym}
\end{figure}

Recall that a half-integer spin and an integer spin transform as projective and linear representations of the $\mathbb D_2$ point group, respectively.
As an example, consider a spin-$\frac12$ located at site $a$. The two generators of $\mathbb D_2$, the $C_2$ rotation and mirror reflection $R_x$, are represented by $U(C_2)=i\sigma_z$ and $U(R_x)=i\sigma_x$ acting on the 2D Hilbert space of a spin-$\frac12$, respectively. These two operators anticommute,
$U(C_2)U(R_x)=-U(R_x)U(C_2)$, which differs from the multiplication rule for group elements in $\mathbb D_2$: $C_2R_x=R_xC_2$. This ``twisted'' relation implies that the 2D Hilbert space of spin-$\frac12$ forms a projective representation of the $\mathbb D_2$ group (see Sec. I of the SM for a brief review of projective representations).

It is important to note that states in the Hilbert space must either all form linear representations or all form projective representations of the same class, because excitations that connect the ground state to excited states all carry linear representations of the symmetry group.

Now consider a quantum spin system on a rectangular lattice with {\it open} boundary condition, which maps onto itself under the point group $G_a$. Such a lattice is translational invariant apart from the boundary.
We ask whether the many-body Hilbert space  $\cal H$ of the system---the direct product of the spin Hilbert space at every site---forms a linear or projective representation of $G_a$. These two cases are denoted by a $\mathbb Z_2$ index $\nu_a =+1$ or $-1$ respectively.
To answer this question, we first note that sites can be grouped into ``orbits'': each orbit consists of those sites that map onto each other under the symmetry operations of $G_a$. For example, any site {\it not} on either of the two mirror axes passing through $a$ belongs to an orbit of four sites, with one in each quadrant. Any site on a mirror axis, other than $a$, belongs to an orbit of two sites that are related by two-fold rotation. With open boundary condition, $a$ is the only fixed point under $G_a$, hence forms an orbit of its own, $\{ a\}$. Since all orbits except $\{ a \} $ contain an even number of sites,
the many-body Hilbert space of all spins other than the one at $a$, forms a linear representation of $G_a$. Therefore, we conclude that the Hilbert space of the entire spin system forms a projective (linear) representation of $G_a=\mathbb D_2$, if and only if the spin at the center $a$ is half-integer (integer), respectively. This result can be expressed by
\begin{equation}
  \label{eq:nua}
  \nu_a=(-1)^{2S_a},
\end{equation}
where $S_a$, the spin at $a$, is either a half-integer or integer.

When $\cal H$ forms a  projective representation, any Hamiltonian invariant under the point group $G_a$ must have degenerate ground states, simply because projective representations necessarily have dimensions greater than $1$. This mathematical fact can be intuitively understood from the non-commutativeness of the algebra $U(C_2)U(R_x)=-U(R_x)U(C_2)$, which can only be realized by matrices of sizes greater than 1~\cite{Hirano2008}. The ground state degeneracy here is thus protected by the point group $G_a=\mathbb D_2$, which is a subgroup of the wallpaper group $G=\text{p2mm}$.


The above result (\ref{eq:nua}) can be applied to to lattices of increasing sizes, which are invariant under $G_a$.
If the spin at site $a$ is half-integer, the point group symmetry $G_a$ guarantees that the ground state degeneracy persists in the thermodynamic limit, with open boundary condition.

We now discuss the implications of this thermodynamic degeneracy for the ground state of the system. First, if the system
has a unique ground state on a torus, the degeneracy shown above must come from boundary degrees of freedom, implying that the ground state of the system is an SPT state protected by the point group $G_a$. Interestingly, if realized, such an SPT state of half-integer spins cannot belong to the known classification which assumes physical degrees of freedom form linear representation of $G_a$. 

The opposite possibility is that the system also has thermodynamically degenerate ground states on the boundary-less torus. This bulk degeneracy may imply that the ground state is  symmetry breaking or topologically ordered.  
If the system is topologically ordered, the degeneracy shown above for open boundary condition can be the result of fractional point-group symmetry quantum numbers~\cite{YCWangEdge} of anyons, implying a symmetry-enriched topological state.

We have thus ruled out completely featureless ground states for quantum spin systems with half-integer spins on symmetry centers, while leaving alive the possibility of SPT states, symmetry-breaking, and symmetry-enriched topological order.
We now further rule out the possibility of an SPT state for systems which have an odd number of half-integer spins on symmetry centers in each unit cell. This is achieved by putting the system on a torus with an odd number of unit cells~\cite{PoLSM2017}
 (as shown in Sec.II of the SM, such a torus can always be constructed compatible with the wallpaper groups). In this setup, the whole system has in total an odd number of half-integer spins on symmetry centers, which together form a projective representation of the $\mathbb D_2$ symmetry group. Therefore, the ground state degeneracy remains on the torus. 
In this case, our result is still in some aspects stronger than the LSM Theorem and its previous generalizations, because it requires only the crystal symmetry, and does not need time-reversal and spin-rotation symmetries.

This argument can be readily generalized to other centers of symmetry. Each center of symmetry $a$ in the 2D lattice has an associated point group $G_a$, which always has the structure of $\mathbb D_2$ for the wallpaper group p2mm. Therefore, for each $a$, the $\mathbb Z_2$ quantum number $\nu_a$ computed from Eq.~\eqref{eq:nua} determines whether the entire system transforms projectively under $G_a$. Furthermore, if two centers of symmetry $a$ and $b$ are related by crystal symmetries, they must host the same quantum number $\nu_a=\nu_b$. Therefore, in p2mm, there are only four independent quantum numbers, as there are four inequivalent centers of symmetry. Any one of them being $-1$ implies that the ground state must have a degeneracy protected by the wallpaper-group symmetry.

\paragraph{Other wallpaper groups.}

We now generalize our result to other 2D wallpaper groups. Similar to the example of p2mm, we consider a center of symmetry $a$ and the associated point group $G_a$. Using the same argument as in the previous example, one can show that the system has a $G_a$-protected ground state degeneracy, if the degrees of freedom at site $a$ transforms projectively under $G_a$. There are eight different types of point groups in 2D: the cyclic groups $\mathbb C_n$ and the dihedral groups $\mathbb D_n$, where $n=2,3,4,6$. Among the eight possible point groups, only three dihedral groups, $\mathbb D_2$, $\mathbb D_4$ and $\mathbb D_6$, have nontrivial projective representations .
They both have a $\mathbb Z_2$ classification of projective representations: one class of linear representation and one class of nontrivial projective representation, realized by an integer spin and a half-integer spin, respectively.
Therefore, for each center of symmetry $a$ with $G_a=\mathbb D_{2,4,6}$, the quantum number $\nu_a$ defined in Eq.~\eqref{eq:nua} reflects whether the entire system transforms projectively under $G_a$. (These point groups all contain a two-fold rotation, which implies that all sites except the center of symmetry form orbits of even sizes.) The number of independent quantum numbers is equal to the number of inequivalent centers of symmetry with $G_a=\mathbb D_{2,4,6}$; any $\nu_a=-1$ implies that the ground state must have a degeneracy protected by the wallpaper-group symmetry.

\begin{table}
\begin{tabularx}{.5\textwidth}{CCC}
\hline\hline
Wallpaper group & Unit cell & \# of $\nu_a$\\
\hline
p2mm & \includegraphics{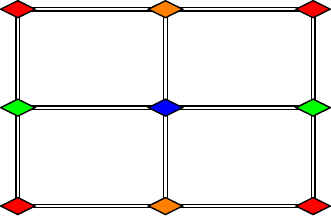} & 4\\
c2mm & \includegraphics{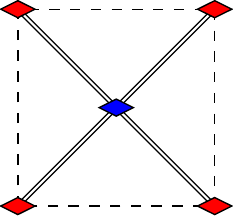} & 2\\
p4mm & \includegraphics{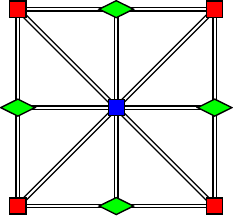} & 3\\
p4gm & \includegraphics{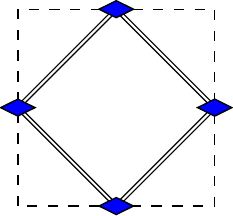} & 1\\
p6mm & \includegraphics{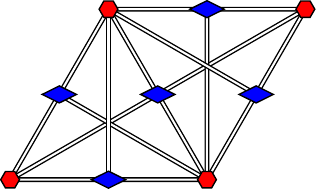} & 2\\
\hline\hline
\end{tabularx}
\caption{List of wallpaper groups where our no-go theorem applies. The figures in the second column illustrates the elements of the wallpaper group: the double lines are mirror axes, and symbols represent centers of symmetry. The rhombuses, squares and hexagons are centers of symmetry with $\mathbb D_2$, $\mathbb D_4$ and $\mathbb D_6$ point groups, respectively. Different colors of the symbols label inequivalent centers of symmetry. The dashed lines outline a unit cell, when it is not bordered by double lines representing mirror axes.}
\label{tab:sgs}
\end{table}

Among the 17 2D wallpaper groups, five of them has centers of symmetry for which a quantum number $\nu_a$ can be defined. We summarize the position of such centers of symmetry and the number of independent $\nu_a$ quantum numbers in Table~\ref{tab:sgs}.
We notice that our no-go theorem does not apply to spin-$\frac12$ models on the honeycomb lattice, because the lattice sites are centers of $\mathbb D_3$ point-group symmetry. Such a center of symmetry cannot be used in our no-go theorem, because $\mathbb D_3$ does not have nontrivial projective representations. This is consistent with the recent construction of a unique symmetric ground state of a honeycomb lattice with spin-$\frac12$ at each lattice site~\cite{PKim2016}.

Similar to the original LSM theorem~\cite{MChengLSMSET2016}, this symmetry-protected ground-state degeneracy can be understood as the surface ground-state degeneracy of a 3D symmetry-protected topological (SPT) state~\cite{XChenSPT2012,XChenSPTCoho2013}. The half-integer-spin degree of freedom that transforms projectively under the point-group symmetry $\mathbb D_{2,4,6}$ can be realized as the edge state of a one-dimensional (1D) Haldane spin chain~\cite{HaldaneChain,HaldaneChainA}, which is a 1D SPT state protected by the point-group symmetry symmetry made of objects that transform linearly~\cite{GuTEF2009,XChen1DSPT2011,XChen1DSPT2011a}. Therefore, the 2D lattice, containing half-integer spins on the centers of symmetry, can be realized as the surface of a 3D system made of a 2D lattice of 1D Haldane chains, which is a 3D SPT state protected by the wallpaper-group symmetry~\cite{HSong2017,IsobeITCI2015}. The surface of such an SPT state must have a symmetry-protected ground-state degeneracy, which is the degeneracy we derived above.

\paragraph{Outlooks.}

\begin{figure}
  \includegraphics{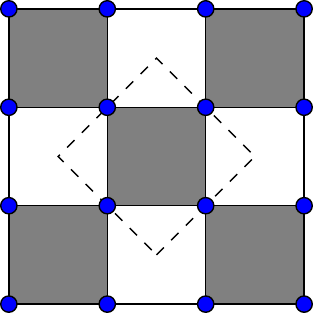}
  \caption{The checkerboard-lattice spin-$\frac12$ model. Each blue circle represents a spin-$\frac12$. The shaded and white plaquettes have different spin-spin interactions. The model has the p4mm symmetry, with a unit cell labeled by the dashed lines.}
  \label{fig:chkbrd}
\end{figure}

Comparing to the LSM Theorem and its recent generalizations, our theorem does not rely on the translation symmetries, and applies to systems with an even number of spin-$\frac12$s per unit cell.
In particular, we consider a spin-$\frac12$ model on a checkerboard lattice. Comparing to a square lattice, a checkerboard lattice symmetry allows different spin-spin interactions on the two types of plaquettes, as shown in Fig.~\ref{fig:chkbrd}. Examples of such models include the checkerboard $J_1$-$J_2$ Heisenberg model~\cite{Singh1998,Oleg2003,LiBishop2015}, which is the effective spin model for the so-called planar-pyrochlore quasi-2D materials~\cite{Kageyama2005,Tsirlin2009,Carretta2009,Skoulatos2009}. The checkerboard lattices has the square-lattice p4mm symmetry, with a unit cell containing two lattice sites (see Fig.~\ref{fig:chkbrd}). Therefore the LSM theorem does not apply. In contrast, our no-go theorem still applies, because the spin-$\frac12$s are located on the $\mathbb D_2$ center of the lattice (marked as green rhombuses in the corresponding row of Table~\ref{tab:sgs}). Consequently, our no-go theorem guarantees a symmetry-protected ground-state degeneracy, indicating that the planar-pyrochlore systems are a promising place to look for topological quantum spin liquids.

When no-go theorems guarantees a ground-state degeneracy, one possibility is that the ground state is a gapped quantum spin liquid state with an intrinsic topological order, which protects a topological ground-state degeneracy, although all local excitations are gapped. In this case, the no-go theorems often put additional constraints on the possible symmetry-fractionalization patterns realized in such spin liquids.
Our no-go theorem can also be extended to provide such constraints: for example, having a projective representation of the $\mathbb D_2$ group at the center of symmetry will determine whether an anyon also carries a projective representation of the $\mathbb D_2$ group~\cite{hsongThesis,QiSpinonPSG}. We leave an extensive study of this extension to a future work. It will be interesting to generalize our result to 3D spin systems, which we also leave to a future work.

\begin{acknowledgements}
We gratefully acknowledge Michael Hermele for very helpful comments on our manuscript.
The work at MIT was supported by DOE Office of Basic Energy Sciences, Division of Materials Sciences and Engineering under Award de-sc0010526.

\emph{Note added.} After completing our manucript we were informed of a related work~\cite{HermeleX}.
\end{acknowledgements}

\bibliography{lsm,checkerboard,hermelex}

\begin{thebibliography}{41}%
\makeatletter
\providecommand \@ifxundefined [1]{%
 \@ifx{#1\undefined}
}%
\providecommand \@ifnum [1]{%
 \ifnum #1\expandafter \@firstoftwo
 \else \expandafter \@secondoftwo
 \fi
}%
\providecommand \@ifx [1]{%
 \ifx #1\expandafter \@firstoftwo
 \else \expandafter \@secondoftwo
 \fi
}%
\providecommand \natexlab [1]{#1}%
\providecommand \enquote  [1]{``#1''}%
\providecommand \bibnamefont  [1]{#1}%
\providecommand \bibfnamefont [1]{#1}%
\providecommand \citenamefont [1]{#1}%
\providecommand \href@noop [0]{\@secondoftwo}%
\providecommand \href [0]{\begingroup \@sanitize@url \@href}%
\providecommand \@href[1]{\@@startlink{#1}\@@href}%
\providecommand \@@href[1]{\endgroup#1\@@endlink}%
\providecommand \@sanitize@url [0]{\catcode `\\12\catcode `\$12\catcode
  `\&12\catcode `\#12\catcode `\^12\catcode `\_12\catcode `\%12\relax}%
\providecommand \@@startlink[1]{}%
\providecommand \@@endlink[0]{}%
\providecommand \url  [0]{\begingroup\@sanitize@url \@url }%
\providecommand \@url [1]{\endgroup\@href {#1}{\urlprefix }}%
\providecommand \urlprefix  [0]{URL }%
\providecommand \Eprint [0]{\href }%
\providecommand \doibase [0]{http://dx.doi.org/}%
\providecommand \selectlanguage [0]{\@gobble}%
\providecommand \bibinfo  [0]{\@secondoftwo}%
\providecommand \bibfield  [0]{\@secondoftwo}%
\providecommand \translation [1]{[#1]}%
\providecommand \BibitemOpen [0]{}%
\providecommand \bibitemStop [0]{}%
\providecommand \bibitemNoStop [0]{.\EOS\space}%
\providecommand \EOS [0]{\spacefactor3000\relax}%
\providecommand \BibitemShut  [1]{\csname bibitem#1\endcsname}%
\let\auto@bib@innerbib\@empty
\bibitem [{\citenamefont {Lieb}\ \emph {et~al.}(1961)\citenamefont {Lieb},
  \citenamefont {Schultz},\ and\ \citenamefont {Mattis}}]{LSM}%
  \BibitemOpen
  \bibfield  {author} {\bibinfo {author} {\bibfnamefont {E.}~\bibnamefont
  {Lieb}}, \bibinfo {author} {\bibfnamefont {T.}~\bibnamefont {Schultz}}, \
  and\ \bibinfo {author} {\bibfnamefont {D.}~\bibnamefont {Mattis}},\ }\href
  {\doibase http://dx.doi.org/10.1016/0003-4916(61)90115-4} {\bibfield
  {journal} {\bibinfo  {journal} {Annals of Physics}\ }\textbf {\bibinfo
  {volume} {16}},\ \bibinfo {pages} {407 } (\bibinfo {year}
  {1961})}\BibitemShut {NoStop}%
\bibitem [{\citenamefont {Affleck}\ and\ \citenamefont
  {Lieb}(1986)}]{Affleck1986}%
  \BibitemOpen
  \bibfield  {author} {\bibinfo {author} {\bibfnamefont {I.}~\bibnamefont
  {Affleck}}\ and\ \bibinfo {author} {\bibfnamefont {E.~H.}\ \bibnamefont
  {Lieb}},\ }\href {\doibase 10.1007/BF00400304} {\bibfield  {journal}
  {\bibinfo  {journal} {Letters in Mathematical Physics}\ }\textbf {\bibinfo
  {volume} {12}},\ \bibinfo {pages} {57} (\bibinfo {year} {1986})}\BibitemShut
  {NoStop}%
\bibitem [{\citenamefont {Affleck}(1988)}]{Affleck1987}%
  \BibitemOpen
  \bibfield  {author} {\bibinfo {author} {\bibfnamefont {I.}~\bibnamefont
  {Affleck}},\ }\href {\doibase 10.1103/PhysRevB.37.5186} {\bibfield  {journal}
  {\bibinfo  {journal} {Phys. Rev. B}\ }\textbf {\bibinfo {volume} {37}},\
  \bibinfo {pages} {5186} (\bibinfo {year} {1988})}\BibitemShut {NoStop}%
\bibitem [{\citenamefont {Yamanaka}\ \emph {et~al.}(1997)\citenamefont
  {Yamanaka}, \citenamefont {Oshikawa},\ and\ \citenamefont
  {Affleck}}]{Yamanaka1997}%
  \BibitemOpen
  \bibfield  {author} {\bibinfo {author} {\bibfnamefont {M.}~\bibnamefont
  {Yamanaka}}, \bibinfo {author} {\bibfnamefont {M.}~\bibnamefont {Oshikawa}},
  \ and\ \bibinfo {author} {\bibfnamefont {I.}~\bibnamefont {Affleck}},\ }\href
  {\doibase 10.1103/PhysRevLett.79.1110} {\bibfield  {journal} {\bibinfo
  {journal} {Phys. Rev. Lett.}\ }\textbf {\bibinfo {volume} {79}},\ \bibinfo
  {pages} {1110} (\bibinfo {year} {1997})}\BibitemShut {NoStop}%
\bibitem [{\citenamefont {Oshikawa}(2000)}]{OshikawaLSM2000}%
  \BibitemOpen
  \bibfield  {author} {\bibinfo {author} {\bibfnamefont {M.}~\bibnamefont
  {Oshikawa}},\ }\href {\doibase 10.1103/PhysRevLett.84.1535} {\bibfield
  {journal} {\bibinfo  {journal} {Phys. Rev. Lett.}\ }\textbf {\bibinfo
  {volume} {84}},\ \bibinfo {pages} {1535} (\bibinfo {year}
  {2000})}\BibitemShut {NoStop}%
\bibitem [{\citenamefont {Hastings}(2004)}]{Hastings2004}%
  \BibitemOpen
  \bibfield  {author} {\bibinfo {author} {\bibfnamefont {M.~B.}\ \bibnamefont
  {Hastings}},\ }\href {\doibase 10.1103/PhysRevB.69.104431} {\bibfield
  {journal} {\bibinfo  {journal} {Phys. Rev. B}\ }\textbf {\bibinfo {volume}
  {69}},\ \bibinfo {pages} {104431} (\bibinfo {year} {2004})}\BibitemShut
  {NoStop}%
\bibitem [{\citenamefont {Balents}(2010)}]{BalentsSLReview}%
  \BibitemOpen
  \bibfield  {author} {\bibinfo {author} {\bibfnamefont {L.}~\bibnamefont
  {Balents}},\ }\href {\doibase 10.1038/nature08917} {\bibfield  {journal}
  {\bibinfo  {journal} {Nature}\ }\textbf {\bibinfo {volume} {464}},\ \bibinfo
  {pages} {199} (\bibinfo {year} {2010})}\BibitemShut {NoStop}%
\bibitem [{\citenamefont {Savary}\ and\ \citenamefont
  {Balents}(2016)}]{SavarySL2016}%
  \BibitemOpen
  \bibfield  {author} {\bibinfo {author} {\bibfnamefont {L.}~\bibnamefont
  {Savary}}\ and\ \bibinfo {author} {\bibfnamefont {L.}~\bibnamefont
  {Balents}},\ }\href {\doibase 10.1088/0034-4885/80/1/016502} {\bibfield
  {journal} {\bibinfo  {journal} {Reports on Progress in Physics}\ }\textbf
  {\bibinfo {volume} {80}},\ \bibinfo {pages} {016502} (\bibinfo {year}
  {2016})}\BibitemShut {NoStop}%
\bibitem [{\citenamefont {Zhou}\ \emph {et~al.}(2017)\citenamefont {Zhou},
  \citenamefont {Kanoda},\ and\ \citenamefont {Ng}}]{ZhouSL2017}%
  \BibitemOpen
  \bibfield  {author} {\bibinfo {author} {\bibfnamefont {Y.}~\bibnamefont
  {Zhou}}, \bibinfo {author} {\bibfnamefont {K.}~\bibnamefont {Kanoda}}, \ and\
  \bibinfo {author} {\bibfnamefont {T.-K.}\ \bibnamefont {Ng}},\ }\href
  {\doibase 10.1103/RevModPhys.89.025003} {\bibfield  {journal} {\bibinfo
  {journal} {Rev. Mod. Phys.}\ }\textbf {\bibinfo {volume} {89}},\ \bibinfo
  {pages} {025003} (\bibinfo {year} {2017})}\BibitemShut {NoStop}%
\bibitem [{\citenamefont {Wen}(1989)}]{WenCSL1989}%
  \BibitemOpen
  \bibfield  {author} {\bibinfo {author} {\bibfnamefont {X.~G.}\ \bibnamefont
  {Wen}},\ }\href {\doibase 10.1103/PhysRevB.40.7387} {\bibfield  {journal}
  {\bibinfo  {journal} {Phys. Rev. B}\ }\textbf {\bibinfo {volume} {40}},\
  \bibinfo {pages} {7387} (\bibinfo {year} {1989})}\BibitemShut {NoStop}%
\bibitem [{\citenamefont {Wen}(1990)}]{WenTO1990}%
  \BibitemOpen
  \bibfield  {author} {\bibinfo {author} {\bibfnamefont {X.~G.}\ \bibnamefont
  {Wen}},\ }\href {\doibase 10.1142/S0217979290000139} {\bibfield  {journal}
  {\bibinfo  {journal} {Int. J. Mod. Phys. B}\ }\textbf {\bibinfo {volume}
  {04}},\ \bibinfo {pages} {239} (\bibinfo {year} {1990})}\BibitemShut
  {NoStop}%
\bibitem [{\citenamefont {Parameswaran}\ \emph {et~al.}(2013)\citenamefont
  {Parameswaran}, \citenamefont {Turner}, \citenamefont {Arovas},\ and\
  \citenamefont {Vishwanath}}]{Parameswaran2013}%
  \BibitemOpen
  \bibfield  {author} {\bibinfo {author} {\bibfnamefont {S.~A.}\ \bibnamefont
  {Parameswaran}}, \bibinfo {author} {\bibfnamefont {A.~M.}\ \bibnamefont
  {Turner}}, \bibinfo {author} {\bibfnamefont {D.~P.}\ \bibnamefont {Arovas}},
  \ and\ \bibinfo {author} {\bibfnamefont {A.}~\bibnamefont {Vishwanath}},\
  }\href {\doibase 10.1038/nphys2600} {\bibfield  {journal} {\bibinfo
  {journal} {Nat. Phys.}\ }\textbf {\bibinfo {volume} {9}},\ \bibinfo {pages}
  {299} (\bibinfo {year} {2013})}\BibitemShut {NoStop}%
\bibitem [{\citenamefont {Roy}()}]{Roy2012X}%
  \BibitemOpen
  \bibfield  {author} {\bibinfo {author} {\bibfnamefont {R.}~\bibnamefont
  {Roy}},\ }\href@noop {} {\ }\Eprint {http://arxiv.org/abs/1212.2944}
  {arXiv:1212.2944 [cond-mat.str-el]} \BibitemShut {NoStop}%
\bibitem [{\citenamefont {Watanabe}\ \emph {et~al.}(2015)\citenamefont
  {Watanabe}, \citenamefont {Po}, \citenamefont {Vishwanath},\ and\
  \citenamefont {Zaletel}}]{Watanabe2015}%
  \BibitemOpen
  \bibfield  {author} {\bibinfo {author} {\bibfnamefont {H.}~\bibnamefont
  {Watanabe}}, \bibinfo {author} {\bibfnamefont {H.~C.}\ \bibnamefont {Po}},
  \bibinfo {author} {\bibfnamefont {A.}~\bibnamefont {Vishwanath}}, \ and\
  \bibinfo {author} {\bibfnamefont {M.}~\bibnamefont {Zaletel}},\ }\href
  {\doibase 10.1073/pnas.1514665112} {\bibfield  {journal} {\bibinfo  {journal}
  {Proc. Natl. Acad. Sci. U.S.A.}\ }\textbf {\bibinfo {volume} {112}},\
  \bibinfo {pages} {14551} (\bibinfo {year} {2015})}\BibitemShut {NoStop}%
\bibitem [{\citenamefont {Po}\ \emph {et~al.}()\citenamefont {Po},
  \citenamefont {Watanabe}, \citenamefont {Jian},\ and\ \citenamefont
  {Zaletel}}]{PoLSM2017}%
  \BibitemOpen
  \bibfield  {author} {\bibinfo {author} {\bibfnamefont {H.~C.}\ \bibnamefont
  {Po}}, \bibinfo {author} {\bibfnamefont {H.}~\bibnamefont {Watanabe}},
  \bibinfo {author} {\bibfnamefont {C.-M.}\ \bibnamefont {Jian}}, \ and\
  \bibinfo {author} {\bibfnamefont {M.~P.}\ \bibnamefont {Zaletel}},\
  }\href@noop {} {\ }\Eprint {http://arxiv.org/abs/1703.06882}
  {arXiv:1703.06882 [cond-mat.str-el]} \BibitemShut {NoStop}%
\bibitem [{\citenamefont {Lu}()}]{YMLuLSM}%
  \BibitemOpen
  \bibfield  {author} {\bibinfo {author} {\bibfnamefont {Y.-M.}\ \bibnamefont
  {Lu}},\ }\href@noop {} {\ }\Eprint {http://arxiv.org/abs/1705.04691}
  {arXiv:1705.04691 [cond-mat.str-el]} \BibitemShut {NoStop}%
\bibitem [{\citenamefont {Yang}\ \emph {et~al.}()\citenamefont {Yang},
  \citenamefont {Jiang}, \citenamefont {Vishwanath},\ and\ \citenamefont
  {Ran}}]{YangDyonLSM}%
  \BibitemOpen
  \bibfield  {author} {\bibinfo {author} {\bibfnamefont {X.}~\bibnamefont
  {Yang}}, \bibinfo {author} {\bibfnamefont {S.}~\bibnamefont {Jiang}},
  \bibinfo {author} {\bibfnamefont {A.}~\bibnamefont {Vishwanath}}, \ and\
  \bibinfo {author} {\bibfnamefont {Y.}~\bibnamefont {Ran}},\ }\href@noop {} {\
  }\Eprint {http://arxiv.org/abs/1705.05421} {arXiv:1705.05421
  [cond-mat.str-el]} \BibitemShut {NoStop}%
\bibitem [{Note1()}]{Note1}%
  \BibitemOpen
  \bibinfo {note} {For simplicity, in this paper we consider a strictly 2D
  lattice with a wallpaper group symmetry, and do not distinguish the symmetry
  groups $\protect \mathbb D_n$ and $C_{nv}$. In general, our theorem applies
  to both cases.}\BibitemShut {Stop}%
\bibitem [{\citenamefont {Hirano}\ \emph {et~al.}(2008)\citenamefont {Hirano},
  \citenamefont {Katsura},\ and\ \citenamefont {Hatsugai}}]{Hirano2008}%
  \BibitemOpen
  \bibfield  {author} {\bibinfo {author} {\bibfnamefont {T.}~\bibnamefont
  {Hirano}}, \bibinfo {author} {\bibfnamefont {H.}~\bibnamefont {Katsura}}, \
  and\ \bibinfo {author} {\bibfnamefont {Y.}~\bibnamefont {Hatsugai}},\ }\href
  {\doibase 10.1103/PhysRevB.78.054431} {\bibfield  {journal} {\bibinfo
  {journal} {Phys. Rev. B}\ }\textbf {\bibinfo {volume} {78}},\ \bibinfo
  {pages} {054431} (\bibinfo {year} {2008})}\BibitemShut {NoStop}%
\bibitem [{\citenamefont {Wang}\ \emph {et~al.}()\citenamefont {Wang},
  \citenamefont {Fang}, \citenamefont {Cheng}, \citenamefont {Qi},\ and\
  \citenamefont {Meng}}]{YCWangEdge}%
  \BibitemOpen
  \bibfield  {author} {\bibinfo {author} {\bibfnamefont {Y.-C.}\ \bibnamefont
  {Wang}}, \bibinfo {author} {\bibfnamefont {C.}~\bibnamefont {Fang}}, \bibinfo
  {author} {\bibfnamefont {M.}~\bibnamefont {Cheng}}, \bibinfo {author}
  {\bibfnamefont {Y.}~\bibnamefont {Qi}}, \ and\ \bibinfo {author}
  {\bibfnamefont {Z.~Y.}\ \bibnamefont {Meng}},\ }\href@noop {} {\ }\Eprint
  {http://arxiv.org/abs/1701.01552} {arXiv:1701.01552 [cond-mat.str-el]}
  \BibitemShut {NoStop}%
\bibitem [{\citenamefont {Kim}\ \emph {et~al.}(2016)\citenamefont {Kim},
  \citenamefont {Lee}, \citenamefont {Jiang}, \citenamefont {Ware},
  \citenamefont {Jian}, \citenamefont {Zaletel}, \citenamefont {Han},\ and\
  \citenamefont {Ran}}]{PKim2016}%
  \BibitemOpen
  \bibfield  {author} {\bibinfo {author} {\bibfnamefont {P.}~\bibnamefont
  {Kim}}, \bibinfo {author} {\bibfnamefont {H.}~\bibnamefont {Lee}}, \bibinfo
  {author} {\bibfnamefont {S.}~\bibnamefont {Jiang}}, \bibinfo {author}
  {\bibfnamefont {B.}~\bibnamefont {Ware}}, \bibinfo {author} {\bibfnamefont
  {C.-M.}\ \bibnamefont {Jian}}, \bibinfo {author} {\bibfnamefont
  {M.}~\bibnamefont {Zaletel}}, \bibinfo {author} {\bibfnamefont {J.~H.}\
  \bibnamefont {Han}}, \ and\ \bibinfo {author} {\bibfnamefont
  {Y.}~\bibnamefont {Ran}},\ }\href {\doibase 10.1103/PhysRevB.94.064432}
  {\bibfield  {journal} {\bibinfo  {journal} {Phys. Rev. B}\ }\textbf {\bibinfo
  {volume} {94}},\ \bibinfo {pages} {064432} (\bibinfo {year}
  {2016})}\BibitemShut {NoStop}%
\bibitem [{\citenamefont {Cheng}\ \emph {et~al.}(2016)\citenamefont {Cheng},
  \citenamefont {Zaletel}, \citenamefont {Barkeshli}, \citenamefont
  {Vishwanath},\ and\ \citenamefont {Bonderson}}]{MChengLSMSET2016}%
  \BibitemOpen
  \bibfield  {author} {\bibinfo {author} {\bibfnamefont {M.}~\bibnamefont
  {Cheng}}, \bibinfo {author} {\bibfnamefont {M.}~\bibnamefont {Zaletel}},
  \bibinfo {author} {\bibfnamefont {M.}~\bibnamefont {Barkeshli}}, \bibinfo
  {author} {\bibfnamefont {A.}~\bibnamefont {Vishwanath}}, \ and\ \bibinfo
  {author} {\bibfnamefont {P.}~\bibnamefont {Bonderson}},\ }\href {\doibase
  10.1103/PhysRevX.6.041068} {\bibfield  {journal} {\bibinfo  {journal} {Phys.
  Rev. X}\ }\textbf {\bibinfo {volume} {6}},\ \bibinfo {pages} {041068}
  (\bibinfo {year} {2016})}\BibitemShut {NoStop}%
\bibitem [{\citenamefont {Chen}\ \emph {et~al.}(2012)\citenamefont {Chen},
  \citenamefont {Gu}, \citenamefont {Liu},\ and\ \citenamefont
  {Wen}}]{XChenSPT2012}%
  \BibitemOpen
  \bibfield  {author} {\bibinfo {author} {\bibfnamefont {X.}~\bibnamefont
  {Chen}}, \bibinfo {author} {\bibfnamefont {Z.-C.}\ \bibnamefont {Gu}},
  \bibinfo {author} {\bibfnamefont {Z.-X.}\ \bibnamefont {Liu}}, \ and\
  \bibinfo {author} {\bibfnamefont {X.-G.}\ \bibnamefont {Wen}},\ }\href
  {\doibase 10.1126/science.1227224} {\bibfield  {journal} {\bibinfo  {journal}
  {Science}\ }\textbf {\bibinfo {volume} {338}},\ \bibinfo {pages} {1604}
  (\bibinfo {year} {2012})}\BibitemShut {NoStop}%
\bibitem [{\citenamefont {Chen}\ \emph {et~al.}(2013)\citenamefont {Chen},
  \citenamefont {Gu}, \citenamefont {Liu},\ and\ \citenamefont
  {Wen}}]{XChenSPTCoho2013}%
  \BibitemOpen
  \bibfield  {author} {\bibinfo {author} {\bibfnamefont {X.}~\bibnamefont
  {Chen}}, \bibinfo {author} {\bibfnamefont {Z.-C.}\ \bibnamefont {Gu}},
  \bibinfo {author} {\bibfnamefont {Z.-X.}\ \bibnamefont {Liu}}, \ and\
  \bibinfo {author} {\bibfnamefont {X.-G.}\ \bibnamefont {Wen}},\ }\href
  {\doibase 10.1103/PhysRevB.87.155114} {\bibfield  {journal} {\bibinfo
  {journal} {Phys. Rev. B}\ }\textbf {\bibinfo {volume} {87}},\ \bibinfo
  {pages} {155114} (\bibinfo {year} {2013})}\BibitemShut {NoStop}%
\bibitem [{\citenamefont {Haldane}(1983{\natexlab{a}})}]{HaldaneChain}%
  \BibitemOpen
  \bibfield  {author} {\bibinfo {author} {\bibfnamefont {F.~D.~M.}\
  \bibnamefont {Haldane}},\ }\href {\doibase 10.1103/PhysRevLett.50.1153}
  {\bibfield  {journal} {\bibinfo  {journal} {Phys. Rev. Lett.}\ }\textbf
  {\bibinfo {volume} {50}},\ \bibinfo {pages} {1153} (\bibinfo {year}
  {1983}{\natexlab{a}})}\BibitemShut {NoStop}%
\bibitem [{\citenamefont {Haldane}(1983{\natexlab{b}})}]{HaldaneChainA}%
  \BibitemOpen
  \bibfield  {author} {\bibinfo {author} {\bibfnamefont {F.~D.~M.}\
  \bibnamefont {Haldane}},\ }\href {\doibase 10.1016/0375-9601(83)90631-X}
  {\bibfield  {journal} {\bibinfo  {journal} {Phys. Lett. A}\ }\textbf
  {\bibinfo {volume} {93}},\ \bibinfo {pages} {464 } (\bibinfo {year}
  {1983}{\natexlab{b}})}\BibitemShut {NoStop}%
\bibitem [{\citenamefont {Gu}\ and\ \citenamefont {Wen}(2009)}]{GuTEF2009}%
  \BibitemOpen
  \bibfield  {author} {\bibinfo {author} {\bibfnamefont {Z.-C.}\ \bibnamefont
  {Gu}}\ and\ \bibinfo {author} {\bibfnamefont {X.-G.}\ \bibnamefont {Wen}},\
  }\href {\doibase 10.1103/PhysRevB.80.155131} {\bibfield  {journal} {\bibinfo
  {journal} {Phys. Rev. B}\ }\textbf {\bibinfo {volume} {80}},\ \bibinfo
  {pages} {155131} (\bibinfo {year} {2009})}\BibitemShut {NoStop}%
\bibitem [{\citenamefont {Chen}\ \emph
  {et~al.}(2011{\natexlab{a}})\citenamefont {Chen}, \citenamefont {Gu},\ and\
  \citenamefont {Wen}}]{XChen1DSPT2011}%
  \BibitemOpen
  \bibfield  {author} {\bibinfo {author} {\bibfnamefont {X.}~\bibnamefont
  {Chen}}, \bibinfo {author} {\bibfnamefont {Z.-C.}\ \bibnamefont {Gu}}, \ and\
  \bibinfo {author} {\bibfnamefont {X.-G.}\ \bibnamefont {Wen}},\ }\href
  {\doibase 10.1103/PhysRevB.83.035107} {\bibfield  {journal} {\bibinfo
  {journal} {Phys. Rev. B}\ }\textbf {\bibinfo {volume} {83}},\ \bibinfo
  {pages} {035107} (\bibinfo {year} {2011}{\natexlab{a}})}\BibitemShut
  {NoStop}%
\bibitem [{\citenamefont {Chen}\ \emph
  {et~al.}(2011{\natexlab{b}})\citenamefont {Chen}, \citenamefont {Gu},\ and\
  \citenamefont {Wen}}]{XChen1DSPT2011a}%
  \BibitemOpen
  \bibfield  {author} {\bibinfo {author} {\bibfnamefont {X.}~\bibnamefont
  {Chen}}, \bibinfo {author} {\bibfnamefont {Z.-C.}\ \bibnamefont {Gu}}, \ and\
  \bibinfo {author} {\bibfnamefont {X.-G.}\ \bibnamefont {Wen}},\ }\href
  {\doibase 10.1103/PhysRevB.84.235128} {\bibfield  {journal} {\bibinfo
  {journal} {Phys. Rev. B}\ }\textbf {\bibinfo {volume} {84}},\ \bibinfo
  {pages} {235128} (\bibinfo {year} {2011}{\natexlab{b}})}\BibitemShut
  {NoStop}%
\bibitem [{\citenamefont {Song}\ \emph {et~al.}(2017)\citenamefont {Song},
  \citenamefont {Huang}, \citenamefont {Fu},\ and\ \citenamefont
  {Hermele}}]{HSong2017}%
  \BibitemOpen
  \bibfield  {author} {\bibinfo {author} {\bibfnamefont {H.}~\bibnamefont
  {Song}}, \bibinfo {author} {\bibfnamefont {S.-J.}\ \bibnamefont {Huang}},
  \bibinfo {author} {\bibfnamefont {L.}~\bibnamefont {Fu}}, \ and\ \bibinfo
  {author} {\bibfnamefont {M.}~\bibnamefont {Hermele}},\ }\href {\doibase
  10.1103/PhysRevX.7.011020} {\bibfield  {journal} {\bibinfo  {journal} {Phys.
  Rev. X}\ }\textbf {\bibinfo {volume} {7}},\ \bibinfo {pages} {011020}
  (\bibinfo {year} {2017})}\BibitemShut {NoStop}%
\bibitem [{\citenamefont {Isobe}\ and\ \citenamefont
  {Fu}(2015)}]{IsobeITCI2015}%
  \BibitemOpen
  \bibfield  {author} {\bibinfo {author} {\bibfnamefont {H.}~\bibnamefont
  {Isobe}}\ and\ \bibinfo {author} {\bibfnamefont {L.}~\bibnamefont {Fu}},\
  }\href {\doibase 10.1103/PhysRevB.92.081304} {\bibfield  {journal} {\bibinfo
  {journal} {Phys. Rev. B}\ }\textbf {\bibinfo {volume} {92}},\ \bibinfo
  {pages} {081304} (\bibinfo {year} {2015})}\BibitemShut {NoStop}%
\bibitem [{\citenamefont {Singh}\ \emph {et~al.}(1998)\citenamefont {Singh},
  \citenamefont {Starykh},\ and\ \citenamefont {Freitas}}]{Singh1998}%
  \BibitemOpen
  \bibfield  {author} {\bibinfo {author} {\bibfnamefont {R.~R.~P.}\
  \bibnamefont {Singh}}, \bibinfo {author} {\bibfnamefont {O.~A.}\ \bibnamefont
  {Starykh}}, \ and\ \bibinfo {author} {\bibfnamefont {P.~J.}\ \bibnamefont
  {Freitas}},\ }\href {\doibase 10.1063/1.367682} {\bibfield  {journal}
  {\bibinfo  {journal} {J. Appl. Phys.}\ }\textbf {\bibinfo {volume} {83}},\
  \bibinfo {pages} {7387} (\bibinfo {year} {1998})}\BibitemShut {NoStop}%
\bibitem [{\citenamefont {Tchernyshyov}\ \emph {et~al.}(2003)\citenamefont
  {Tchernyshyov}, \citenamefont {Starykh}, \citenamefont {Moessner},\ and\
  \citenamefont {Abanov}}]{Oleg2003}%
  \BibitemOpen
  \bibfield  {author} {\bibinfo {author} {\bibfnamefont {O.}~\bibnamefont
  {Tchernyshyov}}, \bibinfo {author} {\bibfnamefont {O.~A.}\ \bibnamefont
  {Starykh}}, \bibinfo {author} {\bibfnamefont {R.}~\bibnamefont {Moessner}}, \
  and\ \bibinfo {author} {\bibfnamefont {A.~G.}\ \bibnamefont {Abanov}},\
  }\href {\doibase 10.1103/PhysRevB.68.144422} {\bibfield  {journal} {\bibinfo
  {journal} {Phys. Rev. B}\ }\textbf {\bibinfo {volume} {68}},\ \bibinfo
  {pages} {144422} (\bibinfo {year} {2003})}\BibitemShut {NoStop}%
\bibitem [{\citenamefont {Li}\ and\ \citenamefont
  {Bishop}(2015)}]{LiBishop2015}%
  \BibitemOpen
  \bibfield  {author} {\bibinfo {author} {\bibfnamefont {P.~H.~Y.}\
  \bibnamefont {Li}}\ and\ \bibinfo {author} {\bibfnamefont {R.~F.}\
  \bibnamefont {Bishop}},\ }\href {\doibase 10.1088/0953-8984/27/38/386002}
  {\bibfield  {journal} {\bibinfo  {journal} {J. Phys. Condens. Matter}\
  }\textbf {\bibinfo {volume} {27}},\ \bibinfo {pages} {386002} (\bibinfo
  {year} {2015})}\BibitemShut {NoStop}%
\bibitem [{\citenamefont {Kageyama}\ \emph {et~al.}(2005)\citenamefont
  {Kageyama}, \citenamefont {Kitano}, \citenamefont {Oba}, \citenamefont
  {Nishi}, \citenamefont {Nagai}, \citenamefont {Hirota}, \citenamefont
  {Viciu}, \citenamefont {Wiley}, \citenamefont {Yasuda}, \citenamefont {Baba},
  \citenamefont {Ajiro},\ and\ \citenamefont {Yoshimura}}]{Kageyama2005}%
  \BibitemOpen
  \bibfield  {author} {\bibinfo {author} {\bibfnamefont {H.}~\bibnamefont
  {Kageyama}}, \bibinfo {author} {\bibfnamefont {T.}~\bibnamefont {Kitano}},
  \bibinfo {author} {\bibfnamefont {N.}~\bibnamefont {Oba}}, \bibinfo {author}
  {\bibfnamefont {M.}~\bibnamefont {Nishi}}, \bibinfo {author} {\bibfnamefont
  {S.}~\bibnamefont {Nagai}}, \bibinfo {author} {\bibfnamefont
  {K.}~\bibnamefont {Hirota}}, \bibinfo {author} {\bibfnamefont
  {L.}~\bibnamefont {Viciu}}, \bibinfo {author} {\bibfnamefont {J.~B.}\
  \bibnamefont {Wiley}}, \bibinfo {author} {\bibfnamefont {J.}~\bibnamefont
  {Yasuda}}, \bibinfo {author} {\bibfnamefont {Y.}~\bibnamefont {Baba}},
  \bibinfo {author} {\bibfnamefont {Y.}~\bibnamefont {Ajiro}}, \ and\ \bibinfo
  {author} {\bibfnamefont {K.}~\bibnamefont {Yoshimura}},\ }\href {\doibase
  10.1143/JPSJ.74.1702} {\bibfield  {journal} {\bibinfo  {journal} {J. Phys.
  Soc. Jpn.}\ }\textbf {\bibinfo {volume} {74}},\ \bibinfo {pages} {1702}
  (\bibinfo {year} {2005})}\BibitemShut {NoStop}%
\bibitem [{\citenamefont {Tsirlin}\ \emph {et~al.}(2009)\citenamefont
  {Tsirlin}, \citenamefont {Schmidt}, \citenamefont {Skourski}, \citenamefont
  {Nath}, \citenamefont {Geibel},\ and\ \citenamefont {Rosner}}]{Tsirlin2009}%
  \BibitemOpen
  \bibfield  {author} {\bibinfo {author} {\bibfnamefont {A.~A.}\ \bibnamefont
  {Tsirlin}}, \bibinfo {author} {\bibfnamefont {B.}~\bibnamefont {Schmidt}},
  \bibinfo {author} {\bibfnamefont {Y.}~\bibnamefont {Skourski}}, \bibinfo
  {author} {\bibfnamefont {R.}~\bibnamefont {Nath}}, \bibinfo {author}
  {\bibfnamefont {C.}~\bibnamefont {Geibel}}, \ and\ \bibinfo {author}
  {\bibfnamefont {H.}~\bibnamefont {Rosner}},\ }\href {\doibase
  10.1103/PhysRevB.80.132407} {\bibfield  {journal} {\bibinfo  {journal} {Phys.
  Rev. B}\ }\textbf {\bibinfo {volume} {80}},\ \bibinfo {pages} {132407}
  (\bibinfo {year} {2009})}\BibitemShut {NoStop}%
\bibitem [{\citenamefont {Carretta}\ \emph {et~al.}(2009)\citenamefont
  {Carretta}, \citenamefont {Filibian}, \citenamefont {Nath}, \citenamefont
  {Geibel},\ and\ \citenamefont {King}}]{Carretta2009}%
  \BibitemOpen
  \bibfield  {author} {\bibinfo {author} {\bibfnamefont {P.}~\bibnamefont
  {Carretta}}, \bibinfo {author} {\bibfnamefont {M.}~\bibnamefont {Filibian}},
  \bibinfo {author} {\bibfnamefont {R.}~\bibnamefont {Nath}}, \bibinfo {author}
  {\bibfnamefont {C.}~\bibnamefont {Geibel}}, \ and\ \bibinfo {author}
  {\bibfnamefont {P.~J.~C.}\ \bibnamefont {King}},\ }\href {\doibase
  10.1103/PhysRevB.79.224432} {\bibfield  {journal} {\bibinfo  {journal} {Phys.
  Rev. B}\ }\textbf {\bibinfo {volume} {79}},\ \bibinfo {pages} {224432}
  (\bibinfo {year} {2009})}\BibitemShut {NoStop}%
\bibitem [{\citenamefont {Skoulatos}\ \emph {et~al.}(2009)\citenamefont
  {Skoulatos}, \citenamefont {Goff}, \citenamefont {Geibel}, \citenamefont
  {Kaul}, \citenamefont {Nath}, \citenamefont {Shannon}, \citenamefont
  {Schmidt}, \citenamefont {Murani}, \citenamefont {Deen}, \citenamefont
  {Enderle},\ and\ \citenamefont {Wildes}}]{Skoulatos2009}%
  \BibitemOpen
  \bibfield  {author} {\bibinfo {author} {\bibfnamefont {M.}~\bibnamefont
  {Skoulatos}}, \bibinfo {author} {\bibfnamefont {J.~P.}\ \bibnamefont {Goff}},
  \bibinfo {author} {\bibfnamefont {C.}~\bibnamefont {Geibel}}, \bibinfo
  {author} {\bibfnamefont {E.~E.}\ \bibnamefont {Kaul}}, \bibinfo {author}
  {\bibfnamefont {R.}~\bibnamefont {Nath}}, \bibinfo {author} {\bibfnamefont
  {N.}~\bibnamefont {Shannon}}, \bibinfo {author} {\bibfnamefont
  {B.}~\bibnamefont {Schmidt}}, \bibinfo {author} {\bibfnamefont {A.~P.}\
  \bibnamefont {Murani}}, \bibinfo {author} {\bibfnamefont {P.~P.}\
  \bibnamefont {Deen}}, \bibinfo {author} {\bibfnamefont {M.}~\bibnamefont
  {Enderle}}, \ and\ \bibinfo {author} {\bibfnamefont {A.~R.}\ \bibnamefont
  {Wildes}},\ }\href {\doibase 10.1209/0295-5075/88/57005} {\bibfield
  {journal} {\bibinfo  {journal} {EPL}\ }\textbf {\bibinfo {volume} {88}},\
  \bibinfo {pages} {57005} (\bibinfo {year} {2009})}\BibitemShut {NoStop}%
\bibitem [{\citenamefont {Song}(2015)}]{hsongThesis}%
  \BibitemOpen
  \bibfield  {author} {\bibinfo {author} {\bibfnamefont {H.}~\bibnamefont
  {Song}},\ }\emph {\bibinfo {title} {Interplay between Symmetry and
  Topological Order in Quantum Spin Systems}},\ \href
  {http://gradworks.umi.com/37/43/3743730.html} {Ph.D. thesis},\ \bibinfo
  {school} {University of Colorado Boulder} (\bibinfo {year}
  {2015})\BibitemShut {NoStop}%
\bibitem [{\citenamefont {Qi}\ and\ \citenamefont {Cheng}()}]{QiSpinonPSG}%
  \BibitemOpen
  \bibfield  {author} {\bibinfo {author} {\bibfnamefont {Y.}~\bibnamefont
  {Qi}}\ and\ \bibinfo {author} {\bibfnamefont {M.}~\bibnamefont {Cheng}},\
  }\href@noop {} {\ }\Eprint {http://arxiv.org/abs/1606.04544}
  {arXiv:1606.04544 [cond-mat.str-el]} \BibitemShut {NoStop}%
\bibitem [{\citenamefont {Huang}\ \emph {et~al.}()\citenamefont {Huang},
  \citenamefont {Song}, \citenamefont {Huang},\ and\ \citenamefont
  {Hermele}}]{HermeleX}%
  \BibitemOpen
  \bibfield  {author} {\bibinfo {author} {\bibfnamefont {S.-J.}\ \bibnamefont
  {Huang}}, \bibinfo {author} {\bibfnamefont {H.}~\bibnamefont {Song}},
  \bibinfo {author} {\bibfnamefont {Y.-P.}\ \bibnamefont {Huang}}, \ and\
  \bibinfo {author} {\bibfnamefont {M.}~\bibnamefont {Hermele}},\ }\href@noop
  {} {\ }\Eprint {http://arxiv.org/abs/1705.09243} {arXiv:1705.09243
  [cond-mat.str-el]} \BibitemShut {NoStop}%
\end{thebibliography}%

\widetext
\clearpage


\section*{Supplementary material}

# Supplemental Material: Ground state degeneracy in quantum spin systems protected by crystal symmetries

Yang Qi,[1] Chen Fang,[2] and Liang Fu[1]

[1]*Department of physics, Massachusetts Institute of Technology, Cambridge, MA 02139, USA*

[2]*Institute of Physics, Chinese Academy of Sciences, Beijing 100190, China*

## I. PROJECTIVE REPRESENTATIONS

In quantum mechanics, states in the Hilbert space $\mathcal{H}$ form representations of the symmetry group $G$, on which a group element $\mathbf{g}$ act as a linear transformation $\phi(\mathbf{g}) \in GL(\mathcal{H})$. Naively, the map $\phi : G \to GL(\mathcal{H})$ should be a group homomorphism that preserves the group algebra,

$$\phi(\mathbf{g})\phi(\mathbf{h}) = \phi(\mathbf{gh}), \tag{1}$$

where the equal sign means that the two linear transformations, $\phi(\mathbf{g})\phi(\mathbf{h})$ and $\phi(\mathbf{gh})$, are identical. In this work, we assume that $G$ is a unitary symmetry group.

However, the condition in Eq. (1) is too restrictive for a quantum system, because two states $|a\rangle$ and $|a'\rangle = e^{i\theta}|a\rangle$ that differ by a U(1) phase are recongnized as two identical states physically, and therefore the Hilbert space $\mathcal{H}$ should be viewed as a projective space instead of a linear space. Correspondingly, the constraint in Eq. (1) can be relaxed to being satisfied projectively,

$$\phi(\mathbf{g})\phi(\mathbf{h}) = \omega(\mathbf{g}, \mathbf{h})\phi(\mathbf{gh}), \tag{2}$$

where $\omega(\mathbf{g}, \mathbf{h})$ is a U(1) phase factor. A representation $\phi$ that satisfies the relaxed constraint (2) with a nontrivial set of factor (the exact meaning of being nontrivial will be explained later) is called a projective representation; in contrast, a representation satisfying the original constraint (1) is called a linear representation.

Inequivalent projective representations, represented by inequivalent sets of factors $\omega(\mathbf{g}, \mathbf{h})$, are classified by the second group cohomology $\mathcal{H}^2[G, \mathrm{U}(1)]$. Mathematically, a set of factors $\omega(\mathbf{g}, \mathbf{h}) \in \mathrm{U}(1)$ is called a two-cochain, and the second group cohomology $\mathcal{H}^2[G, U(1)]$ classifies two-cochains satisfying a constraint called the cocycle condition, modulo an equivalence relation called the coboundary equivalence.

The cocycle equation is a consequence of the associativity of matrix multiplications: $\phi(\mathbf{g})[\phi(\mathbf{h})\phi(\mathbf{k})] = [\phi(\mathbf{g})\phi(\mathbf{h})]\phi((k))$. Using Eq. (2) to simplify both sides of this equation, we derive the following constraint,

$$\omega(\mathbf{h}, \mathbf{k})\omega(\mathbf{g}, \mathbf{hk}) = \omega(\mathbf{g}, \mathbf{h})\omega(\mathbf{gh}, \mathbf{k}). \tag{3}$$

A two-cocycle $\omega$ satisfying Eq. (3) is called a cocycle, and Eq. (3) is called the cocycle equation.

The coboundary equivalence is the result of sorting projectively equivalent representations into the same equivalence class. Two projective representations $\phi_1$ and $\phi_2$ are projectively equivalent if they differ only by a U(1) phase factor,

$$\phi_1(\mathbf{g}) \sim \phi_2(\mathbf{g}) = \phi_1(\mathbf{g})\theta(\mathbf{g}). \tag{4}$$

As a result, we view the two-cocycles $\omega_{1,2}$, deduced from the representations $\phi_1$ and $\phi_2$, respectively, as equivalent cocycles,

$$\omega_1(\mathbf{g}, \mathbf{h}) \sim \omega_2(\mathbf{g}, \mathbf{h}) = \omega_1(\mathbf{g}, \mathbf{h})\frac{\theta(\mathbf{g})\theta(\mathbf{h})}{\theta(\mathbf{gh})}. \tag{5}$$

This is called a coboundary equivalence, as the additional two-cocycle that appeared on the right-hand-side, $\frac{\theta(\mathbf{g})\theta(\mathbf{h})}{\theta(\mathbf{gh})}$, is called the coboundary of the one-cocycle $\theta$.

Inequivalent projective representations are classified by two-cocycles satisfying Eq. (3), modulo the coboundary equivalence in Eq. (5). Mathematically, this is described by the second group cohomology $\mathcal{H}^2[G, \mathrm{U}(1)]$. In particular, the point groups $\mathbb{D}_{2,4,6}$ have a nontrivial second group cohomology: $\mathcal{H}^2[\mathbb{D}_{2,4,6}, \mathrm{U}(1)] = \mathbb{Z}_2$, whereas all other 2D point groups, $\mathbb{D}_3$ and $\mathbb{C}_{2,3,4,6}$, have trivial second group cohomology.

## II. TORUS GEOMETRIES WITH POINT-GROUP SYMMETRIES

In this section, we demonstrate that the argument of ground-state degeneracy can be extended to a torus, if in each unit cell, an odd number of centers of symmetry with a $\mathbb{D}_{2,4,6}$ point group is occupied by a half-integer spin. Since a torus does not have a boundary, for such systems, the possibility that the ground-state degeneracy comes from the gapless edge states of an SPT state is eliminated.

We first notice that there are tori with arbitrary sizes that preserves the $\mathbb{D}_{2,4,6}$ point groups. In fact, since a torus can be realized as a rectangle, with opposite sides glued together

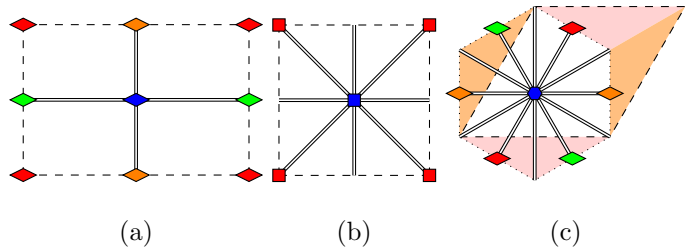

(a) (b) (c)

FIG. 1. Tori preserving point-group symmetries. Here we follow the same convention used in figures of the main text: diamond, square and hexagon symbols label centers of $\mathbb{D}_{2,4,6}$ symmetry, respectively, where symbols marked by the same color are the same point on the torus; double lines mark the mirror axes. The dashed lines mark a rhombus region, which becomes a torus when opposite sides are glued together. a) a rectangular torus preserves the $\mathbb{D}_2$ point-group symmetry, with four centers of symmetry of the same $\mathbb{D}_2$ group. b) a square torus preserves the $\mathbb{D}_4$ point-group symmetry, with two centers of symmetry of the same $\mathbb{D}_4$ group. c) a diamond-shape torus preserves the $\mathbb{D}_6$ point-group symmetry. The diamond-shape region can be cut and glued into a hexagon, marked by a dotted line: the pink and orange shaded regions marks the pieces that need to be cut and glued. There is only one $\mathbb{D}_6$ center of symmetry, but there are three more centers of symmetry, each preserving a $\mathbb{D}_2$ subgroup of $\mathbb{D}_6$.

by periodic boundary conditions, a torus always preserve a $\mathbb{D}_2$ point-group symmetry [see Fig. 1(a)]. Furthermore, a torus with equal number of unit cells on each side also preserves a $\mathbb{D}_4$ point-group symmetry [see Fig. 1(b)]. Lastly, a diamond-shape torus can be cut and glued into a hexagon [see Fig. 1(c)], and therefore preserves a $\mathbb{D}_6$ point-group symmetry. In each case, the torus contains an even number of fixed points, or centers of symmetry, of the point-group symmetry: four for the case of $\mathbb{D}_2$ point group, and two for the cases of $\mathbb{D}_4$. For $\mathbb{D}_6$ symmetry, in additional to the center of symmetry at the center of the hexagon, there are three more centers of symmetry, each fixed under a $\mathbb{D}_2$ subgroup of the whole $\mathbb{D}_6$. If each of them is occupied by a half-integer spin, together they form a projective representation of $\mathbb{D}_6$. Therefore, for all three cases, whether the whole Hilbert space on the torus transforms linearly or projectively is determined by the total spins on all centers of symmetry labeled in Fig. 1. Therefore, the argument presented in the main text cannot always be generalized to a torus without a boundary.

However, such a generalization is possible, if in each unit cell, an odd number of such

centers of symmetry is occupied by half-integer spins. To show this, we consider a torus containing an odd number of unit cells in each direction, and therefore contains an odd number of unit cells in total. Hence, on the whole lattice, there are in total an odd number of centers of $\mathbb{D}_{2,4,6}$ symmetry that are occupied by half-integer spins. As a result, there must be at least one center of symmetry that satisfies the following condition: on the torus, an odd number of fixed points of the associated point group (shown in Fig. 1) is occupied with half-integer spins. Our argument then concludes that the whole Hilbert space on the torus transforms projectively under this point group, and therefore must have symmetry-protected ground-state degeneracy.


\end{document}